\documentclass[apjl]{emulateapj}

\def\mt{}

\def\grb{GRB 090510}

\def\ltsima{$\; \buildrel < \over \sim \;$}
\def\lsim{\lower.5ex\hbox{\ltsima}}
\def\gtsima{$\; \buildrel > \over \sim \;$}
\def\gsim{\lower.5ex\hbox{\gtsima}}

\begin{document}

\bigskip

\author{
A.~Giuliani\altaffilmark{1}, F.~Fuschino\altaffilmark{2}, G.~Vianello\altaffilmark{1},
M.~Marisaldi\altaffilmark{2}, S.~Mereghetti\altaffilmark{1},
M.~Tavani\altaffilmark{3,4}, S.~Cutini\altaffilmark{5},
 G.~Barbiellini\altaffilmark{6}, F.~Longo\altaffilmark{6}, E.~Moretti\altaffilmark{6},
 M.~Feroci\altaffilmark{3}, E.~Del~Monte\altaffilmark{3},
 A.~Argan\altaffilmark{3},
A.~Bulgarelli\altaffilmark{2}, P.~Caraveo\altaffilmark{1},
 P.~W.~Cattaneo\altaffilmark{7},
A.~W.~Chen\altaffilmark{1,14}, T.~Contessi\altaffilmark{1,2},
F.~D'Ammando\altaffilmark{3,4},
E.~Costa\altaffilmark{3}, G.~De Paris\altaffilmark{3},
G.~Di~Cocco\altaffilmark{2}, I.~Donnarumma\altaffilmark{3},
Y.~Evangelista\altaffilmark{3}, A.~Ferrari\altaffilmark{14,18},
M.~Fiorini\altaffilmark{1},
 M.~Galli\altaffilmark{8},
F.~Gianotti\altaffilmark{2}, C.~Labanti\altaffilmark{2},
I.~Lapshov\altaffilmark{3}, F.~Lazzarotto\altaffilmark{3},
P.~Lipari\altaffilmark{9},
A.~Morselli\altaffilmark{11}, L.~Pacciani\altaffilmark{3},
A.~Pellizzoni\altaffilmark{20}, F.~Perotti\altaffilmark{1},
G.~Piano\altaffilmark{3,4,11}, P.~Picozza\altaffilmark{4,11},
M.~Pilia\altaffilmark{12},
G.~Pucella\altaffilmark{13}, M.~Prest\altaffilmark{12},
M.~Rapisarda\altaffilmark{13}, A.~Rappoldi\altaffilmark{7},
A.~Rubini\altaffilmark{3}, S.~Sabatini\altaffilmark{4},
E.~Scalise\altaffilmark{3}, E.~Striani\altaffilmark{3,4,11},
P.~Soffitta\altaffilmark{3}, M.~Trifoglio\altaffilmark{2},
A.~Trois\altaffilmark{3}, E.~Vallazza\altaffilmark{6},
S.~Vercellone\altaffilmark{17}, V.~Vittorini\altaffilmark{3,4},
A.~Zambra\altaffilmark{1}, D.~Zanello\altaffilmark{9},
C.~Pittori\altaffilmark{5},
 F.~Verrecchia\altaffilmark{5}, P.~Santolamazza\altaffilmark{5},
P.~Giommi\altaffilmark{5}, S.~Colafrancesco\altaffilmark{5},
L.A.~Antonelli\altaffilmark{19},  L.~Salotti\altaffilmark{15}}

\altaffiltext{1} {INAF/IASF-Milano, I-20133 Milano, Italy}
\altaffiltext{2} {INAF/IASF-Bologna, I-40129 Bologna, Italy}
\altaffiltext{3} {INAF/IASF-Roma, I-00133 Roma, Italy}
\altaffiltext{4} {Dip. di Fisica, Univ. Tor Vergata, I-00133 Roma, Italy} 
\altaffiltext{5} {ASI Science Data Center, I-00044 Frascati(Roma), Italy} 
\altaffiltext{6} {Dip. Fisica and INFN Trieste, I-34127 Trieste, Italy}
\altaffiltext{7} {INFN-Pavia, I-27100 Pavia, Italy}
\altaffiltext{8} {ENEA-Bologna, I-40129 Bologna, Italy}
\altaffiltext{9} {INFN-Roma La Sapienza, I-00185 Roma, Italy}
\altaffiltext{10} {CNR-IMIP, Roma, Italy} 
\altaffiltext{11} {INFN Roma Tor Vergata, I-00133 Roma, Italy} 
\altaffiltext{12} {Dip. di Fisica, Univ. dell'Insubria, I-22100 Como, Italy}
\altaffiltext{13} {ENEA Frascati,  I-00044 Frascati (Roma), Italy}
\altaffiltext{14} {CIFS-Torino, I-10133 Torino, Italy}
\altaffiltext{15} {Agenzia Spaziale Italiana, I-00198 Roma, Italy} 
\altaffiltext{16} {Osservatorio Astronomico di Trieste, Trieste, Italy} 
\altaffiltext{17}{INAF-IFC, Palermo, Italy}
\altaffiltext{18} {Dip. Fisica, Universit\'a di Torino, Turin, Italy} 
\altaffiltext{19} {INAF-Osservatorio Astron. di Roma, Monte Porzio Catone, Italy}
\altaffiltext{20} {INAF-Osservatorio Astronomico di Cagliari, localita' Poggio dei Pini, strada 54, I-09012 Capoterra, Italy}

\shorttitle{ \grb}
\shortauthors{Giuliani et al. 2009}
\title{ AGILE detection of delayed gamma-ray emission from the
short gamma-ray burst \grb}

\begin{abstract}                                       %
Short gamma-ray bursts (GRBs),  typically lasting less than 2 s,
are a special class of GRBs of great interest. 
We report the detection by the AGILE satellite of the  short GRB 090510 which shows two clearly distinct emission phases: a prompt phase lasting $\sim 200$ msec and a second phase lasting tens of seconds. 
The { prompt phase}  is relatively intense in the 0.3-10 MeV range with a spectrum characterized by a large peak/cutoff energy near 3 MeV,
{ in this phase, no significant high-energy gamma-ray emission is detected}.
{ At the end of the prompt phase, intense gamma-ray emission above 30 MeV is detected showing} 
a power-law time decay of the flux of the type $t^{-1.3}$ and a broad-band spectrum 
remarkably different from that of the prompt phase. 
It extends from sub-MeV to hundreds of MeV energies with a photon index $\alpha \simeq 1.5$. 
{ GRB 090510 provides the first case of a short GRB with delayed gamma-ray emission. }.
We present the timing and spectral data of GRB 090510 and briefly discuss its remarkable properties within the current models of gamma-ray emission of short GRBs.
\end{abstract}                                         %


\maketitle

\section{Introduction}

Gamma-ray bursts (GRBs) are the most energetic explosions in our Universe but only a few bursts were detected at gamma-ray energies above 100 MeV. 
The EGRET instrument on board the {Compton Gamma Ray Observatory} during its 6-year lifetime detected 5 GRBs above 100 MeV \cite{dingus2001}. 
Today, the currently operating AGILE and \textit{Fermi} satellites, have doubled the sample of GRBs detected at these energies 
({e.g., Giuliani et al. 2008, McEnery et al. 2008, Abdo et al.  2009}).
However, the great majority of GRBs with detected photons above 100 MeV  are long bursts with typical durations above 2 seconds: they are possibly associated with stellar explosions of massive stars. 
Much less is known about the high-energy proprieties of {\it short} GRBs that show durations below 2 seconds. 
These short events are usually hard compared to the average properties of GRBs and are believed to be associated with the coalescence of neutron-star binaries { (but see Zhang et al. 2009 for a more thorough discussion of the GRB classification and possible origin of the different classes)}. 
It is then very important to establish the gamma-ray properties of short GRBs.
Before the advent of AGILE and \textit{Fermi} no short-GRB was detected above a few MeV.  
The first short-GRB detection in the gamma-ray energy band was by \textit{Fermi}: GRB~081024B (lasting about 0.8 s in the MeV range)  was detected up to 3 GeV within the first 5 sec after trigger \citep{omodei2008, connaughton}. 
We report here the \textit{AGILE} detection of \grb ,  the second short-GRB detected above 100 MeV.

The Italian AGILE satellite for gamma-ray astronomy has been
operating since 2007 April  \citep{tavani-1}. The Gamma-Ray
Imaging Detector (GRID, Barbiellini et al., 2002) on board AGILE
covers one fifth of the sky in the 30 MeV -- 30 GeV energy range.
This large field of view, together with a gamma-ray detection
deadtime of order of $\sim$100 $\mu$s, makes it particularly
suited for the observation of GRBs. The GRID high-energy data are
complemented by those of other detectors on board the satellite,
which operate in different energy ranges. Super-AGILE provides GRB
localizations, lightcurves and spectra in the 18--60 keV range
\citep{feroci09,Feroci2007,Del_Monte_et_al_2008}. The
Mini-Calorimeter (MCAL), besides being used as part of the GRID,
can be used to autonomously detect and study GRBs in the 0.35--100
MeV range with excellent timing \citep{Labanti2009,Marisaldi2008}.
Finally, GRB lightcurves in the hard X-ray band can be obtained
also from the GRID anti coincidence scintillator panels
\citep{perotti}.

\section{GRB 090510}

The \grb\ was discovered and precisely localized by the \textit{Swift}
satellite \citep{gehrels04} with coordinates (J2000) R.A. = 22h 14m
12.47s, Dec.=--26d 35' 00.4" \citep{GCN_9331} . 
This burst was quite bright, with peak flux $\sim$10 ph cm$^{-2}$ s$^{-1}$ 
{\mt in the energy band 15-150 keV} \cite{GCN_9337}, and was independently
detected also by Konus-Wind \cite{GCN_9344}, Suzaku-WAM \cite{GCN_9335} and \textit{Fermi}-GBM
\cite{GCN_9336}. The main emission lasts about 0.2 s with a
multi-peak structure. Follow-up observations of the optical
transient of \grb\ led to the determination of the redshift $z=
0.903 \pm 0.003 $ \citep{rau}.

This GRB occurred at the border of the standard AGILE-GRID field of view, at an off-axis angle of 61
degrees. 
At this large off-axis angle, the AGILE-GRID effective area is $\sim$100 cm$^{2}$ for photon energies above 25 MeV. 
A quick look analysis of the GRID data showed an excess of photons above 30 MeV consistent with the direction of \grb\ \cite{GCN_9343}. 
\grb\ was also clearly detected in the 0.3-10 MeV energy range with the AGILE-MCAL, while it  was not detected by Super-AGILE, owing to its large off-axis position. 
Emission above a few tens of MeV was also detected by the \textit{Fermi}-LAT instrument \cite{GCN_9334}.

\section{AGILE Timing and Spectral Data}

In the following, we refer all the times to $T_0$ corresponding to 00:23:00.5  UT of 2009, May 10.  
This corresponds to the time of the sharp initial increase of the GRB lightcurve in the MCAL detector. 
Based on the properties of the $0.3-10$ MeV and $\geq 25$ MeV emissions of \grb\, showing a clear
dichotomy between the low- and high-energy gamma-ray emissions, we define two time intervals, 
Interval I from $T_0$ to $T_0+0.20$ s and  Interval II from $T_0+0.20$ s to $T_0+1.20$ s 
(see figure \ref{fig-2}).

\subsection{The prompt phase (Interval I)}

\begin{figure}
\begin{center}
\includegraphics[width=9.cm]{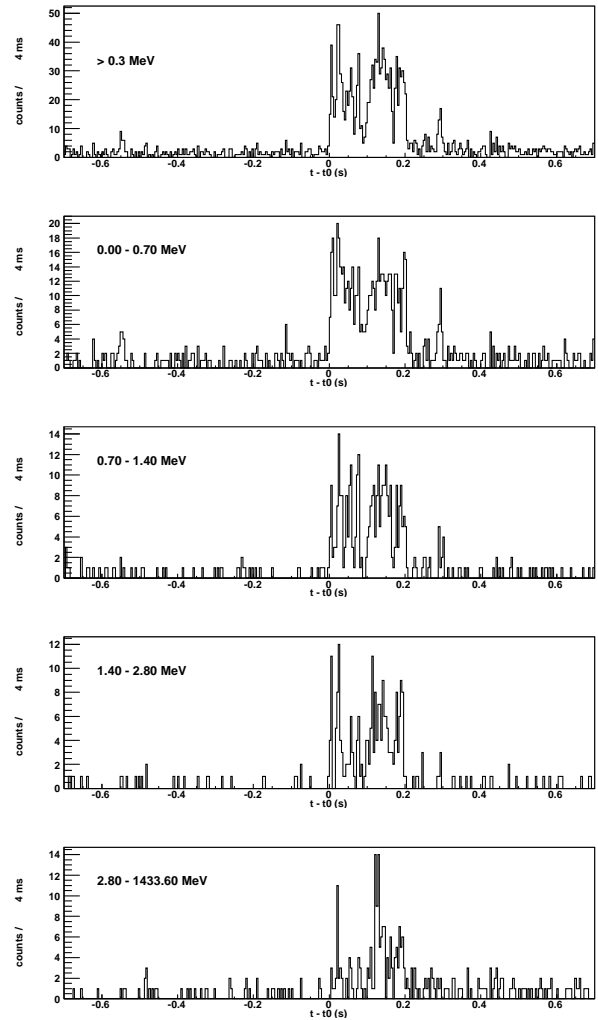}
\caption{Lightcurve of \grb\ as detected by the AGILE-MCAL
detector in different energy ranges.
 The time bin is 4 msec.}
\label{fig-1a}
\end{center}
\end{figure}

\begin{figure}
\begin{center}
\includegraphics[width=9cm, height=11cm]{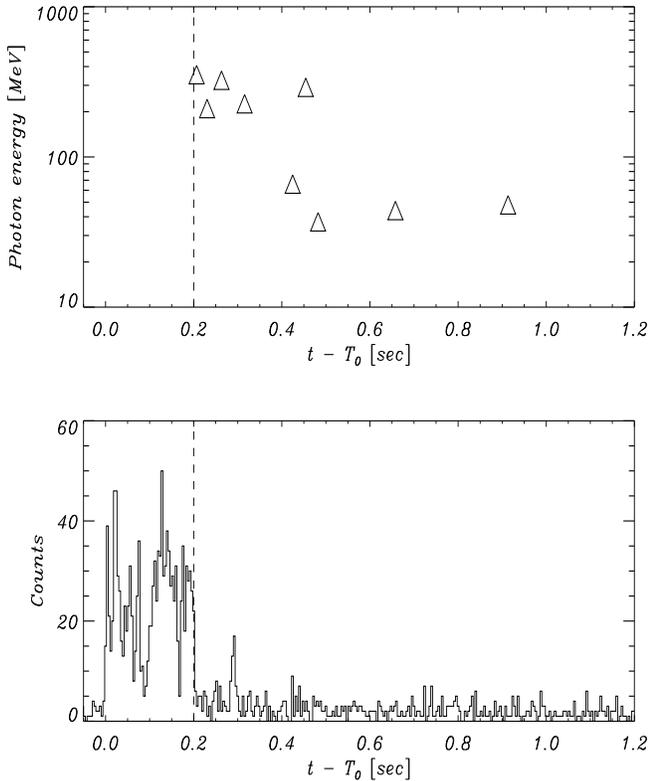}
\caption{(Top panel:) Energies versus arrival time of the GRB  photons  detected  by the AGILE-GRID in the 25 MeV -- 1 GeV energy range. { Note the remarkable absence of gamma-ray events before $T_0 + 0.2$ s}. (Lower panel:)  0.3 -- 10 MeV light curve measured with the AGILE-MCAL detector. The dashed line separates \textit{Interval I} from \textit{Interval II} (see text).}
\label{fig-2}
\end{center}
\end{figure}

The lightcurves of \grb\ obtained with the AGILE-MCAL in the 0.3-10 MeV are shown in Fig.~\ref{fig-1a}. As seen by MCAL the burst has a duration (T90) of $184\pm6$ ms. 
During the T90 time interval MCAL recorded from the source more than 1000 counts above 330~keV, with an expected background of 60 counts over the same time interval. 
The peak flux of 18000~counts/s in a 1~ms time bin was reached at time $T_0 + 0.024$ s. 
To date, this is the brightest short burst detected by MCAL in the GRID field of view. 
In the T90 time interval the observed emission can be divided into three main pulses, each of them showing millisecond time variability. 
{\mt At $T_0 - 0.55$ s a soft  precursor lasting 15~ms is detected up to 700~keV}, 
while at $T_0 + 0.29$ s another 15~ms peak is evident, with significant detection up to
few MeV.

Most of the soft-gamma emission (E$\leq$10 MeV) is concentrated in \textit{Interval I} (between $T_0$ and \textbf{$T_0+0.20$}), where no high-energy (E$\geq$25 MeV) photons were detected.
We derive the flux spectrum for the Interval I using the MCAL  data. %
We find that the averaged MCAL spectrum is well described by a power-law model with exponential cutoff (reduced $\chi^2$ of 0.8 for 23 d.o.f.). The photon index is
 $\alpha_1 = 0.65 (-0.32 + 0.28)$
and the exponential cutoff energy is
$E_{c}= 2.8 (-0.6, +0.9)$ MeV.
 The integrated fluence (500 keV $ \leq E \leq 10$ MeV) during this interval is $F =  1.82 (-0.41, +0.09) \times 10^{-5}$~erg cm$^{-2}$, all the errors for MCAL results reported
throughout this paper are at the 90\% confidence level.
The GRID upper limit (at $3$-$\sigma$ c.l.) is consistent with the extrapolation of this spectrum. 
{\mt  The top panel of Fig.~4 shows the Interval I spectrum.}
{\mt We also notice a substantial soft-to-hard spectral evolution during Interval I. 
If we define a hardness ratio as $HR =$ (counts above 1 MeV)/(counts below 1 MeV), we obtain $HR \sim 0.6 \pm 0.1$ during the first peak (between $T_0$ and $T_0+0.12$) and $HR \sim 0.9 \pm 0.1$ during the second phase of Interval I (between $T_0+0.12$ and $T_0+0.20$).}

A remarkable absence of gamma-ray events during Interval I is evident. 
In fact the first GRID events are detected only at the end of the prompt emission.
Note that a backward extrapolation to $t=T_0+0.01$ s of the GRID power-law lightcurve of Fig.~3, discussed in the next section,  would predict 28 photons in Interval I, while none was observed. 

\subsection{The delayed emission phase: Interval II and tail}

The phase immediately following the sharp decay of the MeV flux at the end of Interval I shows a significant tail of MeV emission and the presence of a strong gamma-ray component above 30 MeV. 
We consider here the Interval II 
and a following tail of emission (lasting up to $T_0+10$ s).

During Interval II and tail, MCAL  continues to detect significant emission in the 500 keV-10 MeV energy range with a spectrum significantly different from that of Interval I. 
Indeed, the derived power-law distribution has now photon index $\alpha_2 =
-1.58 (-0.11, +0.13)$. 
Significant emission is detected in the MCAL highest energy channels with no sign of cutoff.
The Interval II fluence in the energy range 0.5-10 MeV range is $F_2 = 3.1 (-0.7, +0.6) \times 10^{-6} \, \rm erg \, cm^{-2}$.

To search for emission above 30 MeV, we selected GRID gamma-ray events within 15 degrees of the burst position.
For this analysis we used all the GRID events
with reliable direction and energy reconstructions,  resulting in 15 events in the time interval from $T_0$ to $T_0+10$ s. 
The expected number of background events in this time interval is 1.4, implying that the GRB is detected above 30 MeV with a $\geq 5-\sigma$ statistical significance.

\begin{figure}
\begin{center}
\includegraphics[width=9cm]{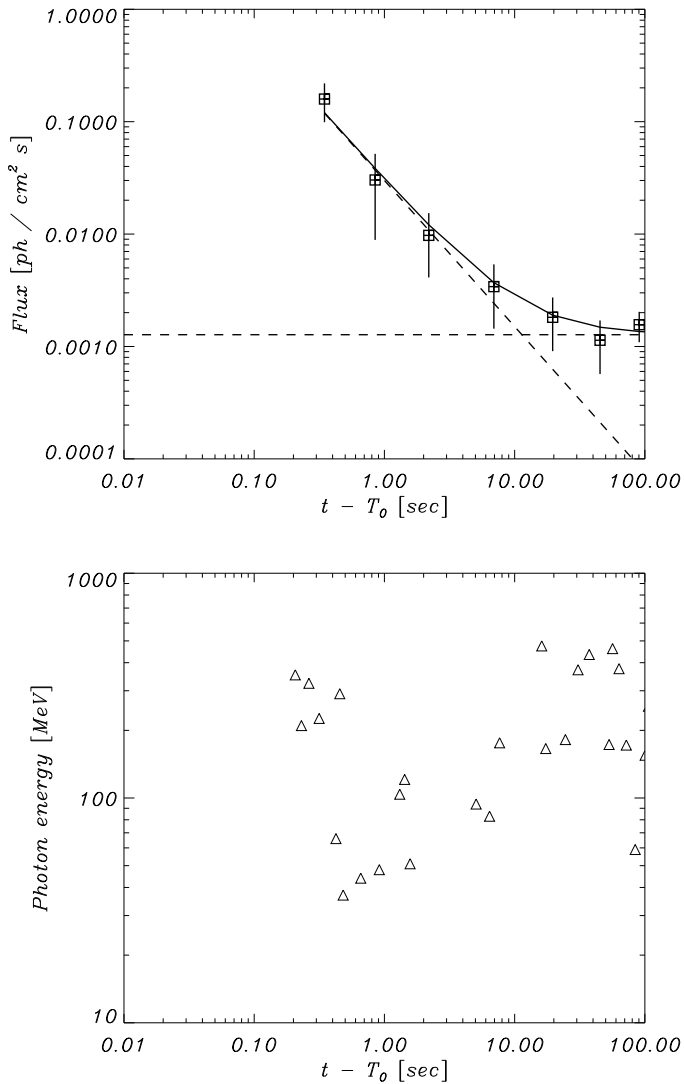}%
\caption{(Top panel:) AGILE-GRID gamma-ray lightcurve of \grb\ for photon events within a sky region of radius 15 degrees. 
The inclined dashed line corresponds to a power-law time decay  $t^{-\delta}$, with $\delta = 1.3$. 
The horizontal dashed line corresponds to the background flux measured in the 1000 seconds before the trigger. The solid line is the sum of the two components.
\\(Bottom panel:) Energies versus arrival time of the GRB  photons  detected  by the AGILE-GRID in the 25 MeV -- 1 GeV energy range. Note that after $T_0+10$ the detected counts are compatible with the background, as shown by the light curve in the top pannel.} 
\vspace{0.5cm}
\label{fig-5}
\end{center}
\end{figure}

The energy and arrival times of the GRID events are compared with the MCAL lightcurve in
Fig. \ref{fig-2}.
The GRID high-energy  emission lasts for a few tens of
seconds after the end of Interval I.
The time evolution of the gamma-ray emission from \grb\ can be remarkably well described
by a power-law decay, as shown in Fig.~\ref{fig-5} (top panel). We
model it with a function given by
\begin{equation}
    F(t) \propto   t^{-\delta} \;\; \mathrm{ for} \;\;  t \geq T_0 + T_1
\end{equation}
and
$ F(t)=0 \;\; \mathrm{for} \;\;  t \leq T_0 + T_1$
and find that $T_1 = 0.2$ s, and $\delta = 1.30 \pm 0.15$ give the
highest probability to reproduce the observed times of arrival.
The corresponding power-law is plotted in the Fig.~3 (top
panel). 
The background flux measured in the 1000 seconds before trigger is also shown in figure by the 
the dashed horizontal line.

The energy distribution of the GRID photons during the time
interval $T_0$ to $T_0+10$ s is consistent with a power-law
spectrum of photon index $\alpha_3 = 1.4 \pm0.4$ ($1$-$\sigma$
c.l.). For this spectrum the 25 MeV- 500 MeV
fluence in the same time interval is $(1.51\pm0.39)\times10^{-1}$
ph cm$^{-2}$, corresponding to $F_3 = (2.90\pm0.75)\times10^{-5}
\rm \, erg \,cm^{-2}$.

\begin{figure}[ht]
\begin{center}
\includegraphics[width=9cm,height=6cm]{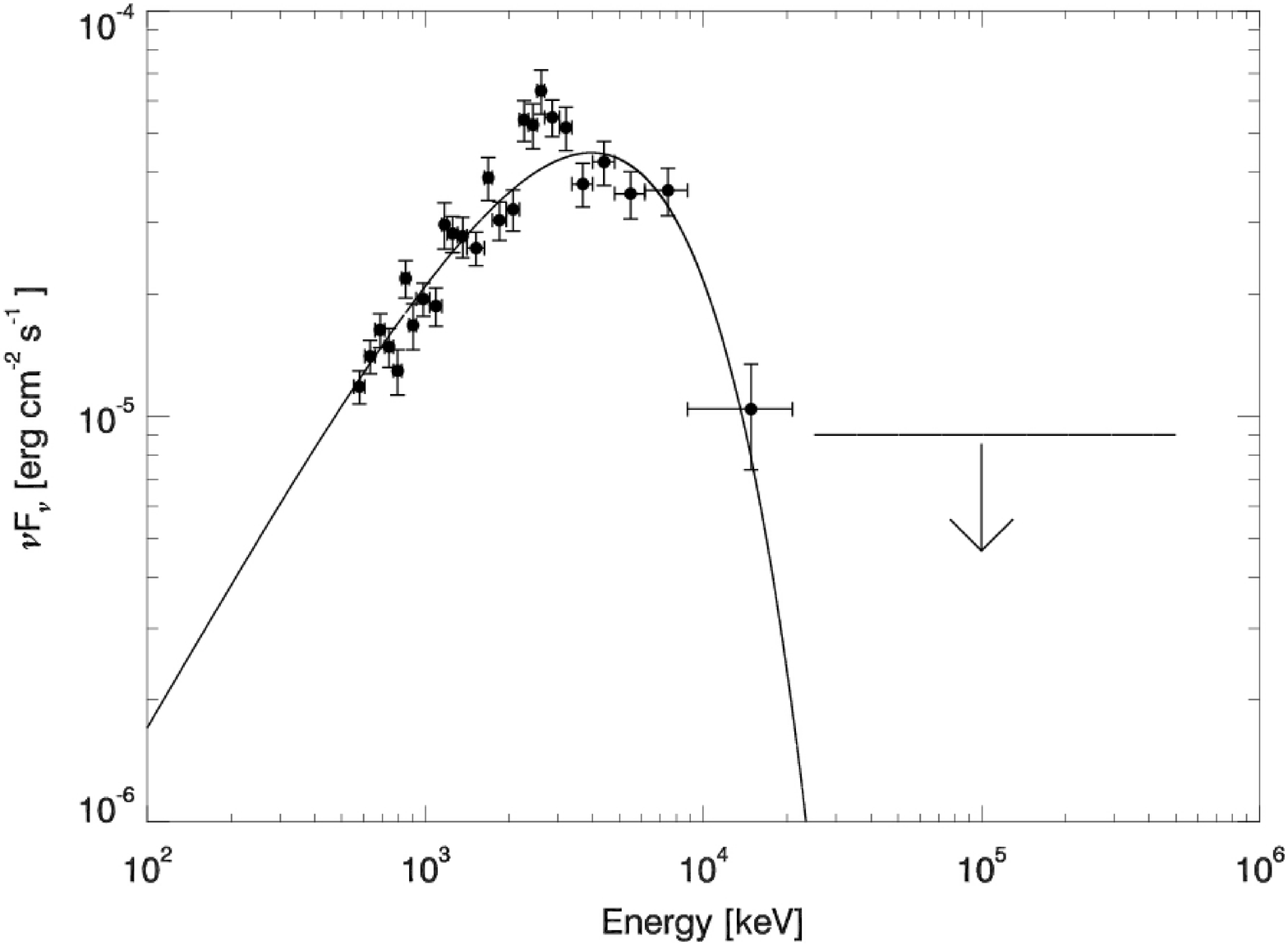}    %
   \includegraphics[width=9cm,height=6cm]{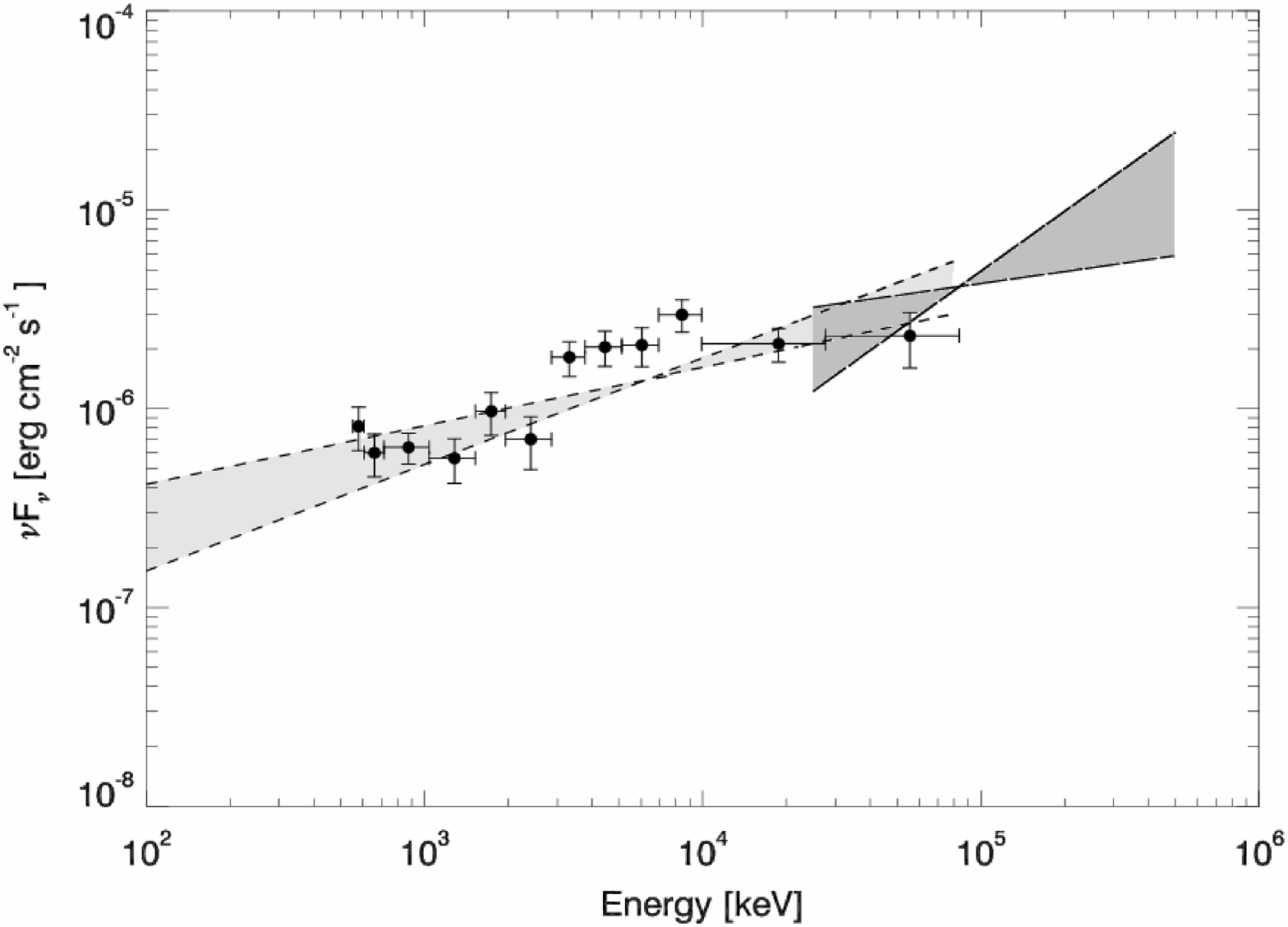}     %
\caption{ (Top panel:) gamma-ray power spectrum of \grb\ for
\textit{Interval I}.  (Bottom panel:) gamma-ray power
spectrum for \textit{Interval II}} \label{spec}
\end{center}
\end{figure}

To compute the GRID flux in Interval II, we assumed the same spectrum measured in the long ''tail" between $T_0$ and $T_0+10$ (photon index $-1.45\pm0.07$), and extrapolated the light curve with the best fit power-law decay. 
This gives $F'_2 = 2.12 \times 10^{-5}$ erg cm$^{-2}$ (25 MeV $ \leq E \leq  500$ MeV).

The MCAL + GRID spectrum of Interval II is shown in the bottom
panel of figure \ref{spec}.

\section{Discussion}

Even though a theoretical investigation is beyond the scope of this
paper, we can briefly emphasize here a few relevant points.

(1) The broad-band emission of \grb\ shows very distinct radiation phases during Interval I and the following delayed emission phase. 
Prompt gamma-rays above 30 MeV are absent during Interval I, but they constitute a crucial component during Interval II and the following tail.
{ Remarkably, this is the first case of a delayed rise of the gamma-ray emission above 25 MeV detected in a short GRB. 
A similar behaviour was shown by the long GRB 080916C \cite{abdo09a}.
This fact suggests that the same process responsible for high energy gamma-ray production takes place, in both long and short GRBs, independently from the central engine.}

(2) The prompt phase (Interval I) spectrum is peaked at $E_p = 3.78$~MeV. 
Comparing  with other short GRBs (see  Ghirlanda et al. 2009) we find that the $E_p$ for GRB 090510 is the highest peak energy ever recorded for a short GRB (about 2.4-$\sigma$ greater than the mean value for short GRBs). Also the rest-frame peak energy ($E_p^{rest} = 7.19$ $(-1.54, +2.31)$~MeV) for \grb is greater than the $E_p^{rest}$ for the other short GRBs with known redshift.
The large value of $E_p^{rest}$, combined with a quite usual value of the 
isotropic (comoving) energetics ($E_{iso,1} = 3.91$ $(-0.88, +1.91)$ $\times 10^{52}$ erg in the whole energy range) and  peak luminosity ($L_{iso,1} = 7.74$ $(-1.74, +3.79)$ $\times 10^{53}$ erg/s)
implies that GRB 090510 does not follow neither the $E_p^{rest}-E_{iso}$ Amati relation (Amati et al 2002) nor the $E_p^{rest}-L_{iso}$ Yonetoku relation (Yonetoku et al. 2004). 
To our knowledge this is the first short GRB that does not follow the Yonetoku relation.
\\No significant emission is detected above 10 MeV, implying a rather strong constraint on any possible
power-law emission above $E_c$ ($\beta < -3.2$  at the  90\% confidence level). 

{\mt (3) The prompt phase shows a significant soft to hard
spectral evolution. As it can be inferred from Fig.~1 and from the MCAL hardness ratio
calculations, the last
peaks of Interval I are harder than the first peak.}

(4) Gamma-ray emission above 25 MeV extends in time for tens of seconds,
i.e.,  well beyond the prompt phase duration, and shows a temporal
behavior consistent with a power-law of index $\delta = 1.30 \pm
0.15$.

(5) The total isotropic energy of Interval II and tail is larger than that of Interval I: 
by summing the MCAL and GRID contributions to the emission, we obtain for the delayed
phase $E_{iso,2} = 4.8$ $\times 10^{52}$ erg.

{\mt (6) The temporal index $\delta = 1.3$ is substantially different from that ($\delta'= 0.75$) subsequently measured by \textit{Swift}-XRT between 80 and 1400 sec after trigger \cite{GCN_9341}. 
This last phase can be attributed to an afterglow with spectral and temporal characteristics in agreement with expectations of fireball models \citep{zhang}. }

High-energy emission from \grb\ can have different physical origins at different phases. 
It is possible to evaluate a lower limit for the Lorentz factor  of the emitting regions
in interval I and II, on the basis of their spectral features and time-scale
variability (Lithwich et al. 2001).  

During Interval I the energy of the highest bin of the spectrum with
significative detection is $E_{max}=20$ MeV. 
A physical scenario with the minimum Lorentz factor compatible with the data 
corresponds to a shell, optically thick for photons of energy greater than $m_e c^2$ (in
the shell rest frame), moving with   $\Gamma_I \geq (1+z)E_{max}/m_e \, c^2 \simeq 80$.
Otherwise, assuming that the emitting region is optically thin also for photons with energy greater than $m_e c^2$,
a larger Lorentz factor is needed ($\Gamma_I \geq 150$), due to the fast variability during this phase,
according with equation 5 in (Lithwich et al. 2001).

Emission during Interval II and following tail appears to be of a very different nature, and is clearly non-thermal. 
Several mechanisms can be at work, depending on the external environment and radiative conditions. Both synchrotron and inverse Compton (IC) emitting regions characterized by impulsively energized
particles can be important contributors. 
The ultimate origin of fast and efficient acceleration is believed to be hydrodynamical
shocks produced by expanding matter ejecta. 
Internal (IS) and external (ES) shocks can in principle contribute to both the synchrotron and IC emissions, and several models have been recently proposed to address the issue of the prompt vs. the so-called "delayed" high-energy emission from GRBs. 
During Interval II the larger photon energy detected is $E_{max}=350$ MeV.
The minimum Lorentz factor compatible with $E_{max}$ is $\Gamma_{II} \geq 200$. 

We postpone an investigation of these issues to forthcoming publications.

\begin{acknowledgements}

The \textit{AGILE} Mission is funded by the Italian Space Agency (ASI) with
scientific and programmatic participation by the Italian Institute
of Astrophysics (INAF) and the Italian Institute of Nuclear
Physics (INFN).
\end{acknowledgements}
\bibliographystyle{aa}

\end{document}